\documentclass{emulateapj}
\usepackage{graphicx,epsfig}

\slugcomment{accepted by the Astrophysical Journal Letters}

\shorttitle{Induced Nested Bars in Assembling Halos}

\begin{document}

\def\gtorder{\mathrel{\raise.3ex\hbox{$>$}\mkern-14mu
     \lower0.6ex\hbox{$\sim$}}}
\def\ltorder{\mathrel{\raise.3ex\hbox{$<$}\mkern-14mu
     \lower0.6ex\hbox{$\sim$}}}

\title{Induced Nested Galactic Bars Inside Assembling Dark Matter Halos
}

\author{
Clayton H.\, Heller,\altaffilmark{1} Isaac\, Shlosman,\altaffilmark{2} 
and E.\, Athanassoula\altaffilmark{3}
}

\altaffiltext{1}{Department of Physics \& Astronomy, Georgia Southern University,
    Statesboro, GA 30460, USA}

\altaffiltext{2}{Department of Physics and Astronomy,
University of Kentucky, Lexington, KY 40506-0055, USA}

\altaffiltext{3}{Observatorie de Marseille, 2 Place Le Verrier, 13248 
    Marseille Cedex 04, France}

\begin{abstract}
We investigate the formation and evolution of nested bar systems in disk galaxies
in a cosmological setting by following the development of an isolated dark matter (DM) and baryon 
density perturbation. The disks form within the 
assembling triaxial DM halos and the feedback from 
the stellar evolution is accounted for in terms of supernovae and OB stellar winds. 
Focusing on a representative model, we show the formation of an oval
disk and of a first generation of nested bars with 
characteristic sub-kpc and a few kpc sizes. The system evolves through successive 
dynamical couplings and decouplings, forcing the gas inwards and
settles in a state of resonant coupling. The inflow rate can support a
broad range of activity within the central kpc, from quasar- to
Seyfert-types, supplemented by a vigorous star formation as a
by-product. The initial bar formation is triggered in response to the
tidal torques from the triaxial DM halo, which acts as a finite
perturbation. This {\it first} generation of bars does not survive
for more than 4--5~Gyr: by that time the secondary bar has totally dissolved, 
while the primary one has very substantially weakened, reduced to a
fat oval. This evolution is largely
due to chaos introduced by the interaction of the multiple non-axisymmetric 
components.  
\end{abstract}
 

%%%%%%%%%%%%%%%%%%%%%%%%%%%%%%%%%%%%%%%%%%%%%%%%%%%%%%%%%%%%%%%%%
\section{Introduction} 
%%%%%%%%%%%%%%%%%%%%%%%%%%%%%%%%%%%%%%%%%%%%%%%%%%%%%%%%%%%%%%%%%%

The formation and evolution of galaxies are closely associated with the radial redistribution 
of baryonic and dark matter (DM) and of their angular momentum ($J$) (e.g., Athanassoula 
2002, 2003). This process is further amplified by the lack of an axial symmetry 
in the basic galactic 
components over substantial periods of time. Numerical simulations have shown
that DM halos form universally triaxial, i.e., flattened and elliptical\footnote{We define
the density (equatorial) ellipticity as $\epsilon_\rho=1-b/a$ --- where $b/a$ is the
intermediate-to-major axis ratio and the flatness as $f_\rho=1-c/a$, 
where $c/a$ --- the minor-to-major axis ratio} (e.g., Allgood et al. 2006), 
and remain in this 
state for the few Gyr during the disk growth (Berentzen \& Shlosman 2006).
A clear majority of disks are barred in the nearby universe (e.g., Knapen et al. 2000; Grosbol et al. 2004;
Marinova \& Jogee 2006), and a fraction of barred disks is maintained
at least to $z\sim 1$ (Jogee et al. 2004; Elmegreen et al. 2004; Sheth et al. 2003).
Simulations have also shown that galactic disks are subject 
to the spontaneous and induced stellar bar formation (e.g., Athanassoula 1984). A wide range of studies, theoretical 
and observational, argue that disk galaxies spend a substantial
fraction of their life in 
a non-axisymmetric stage. Their morphological components are subjected to mutual 
gravitational torques that, in addition to external factors, provide an efficient 
mechanism for driving their internal 
evolution (e.g., Weinberg 1985; Athanassoula 1992;
Heller \& Shlosman 1994; Sellwood 2006).
 
The shapes of the host halos affect strongly the growing disk,
especially because early disks are dominated
by the {\it dissipative} baryonic component, unlike disks in the nearby universe.
The gas responds dramatically to a non-axisymmetric driving by shocking
and loses its rotational support. One can expect that bars or grand-design
spiral arms will be triggered in the disk and that the central mass concentration will build up.
The most efficient redistribution of $J$, down to the smallest radii, 
in a disk can be attributed to the mechanism of nested bars, stellar and gaseous, the
so-called {\it bars-in-bars} scenario (Shlosman et al. 1989, 1990). The system of nested bars 
that consists of the primary (large-scale) and secondary (sub-kpc) bars tumbling with 
different pattern speeds, can facilitate the radial inflow of gas and fuel the 
accretion processes onto supermassive black holes (SBHs). The latter
correlate tightly with the properties of galactic bulges across seven decades in radius 
at least (e.g., Ferrarese \& Ford 2005). About 1/3 of barred galaxies have 
secondary bars (Laine et al. 2002; Erwin \& Sparke 2002). Furthermore, 
formation of the SBH can be triggered by this process, because it leads to a 
specific entropy minimum in the center (Begelman et al. 2006). The dynamics of nested bars
has been investigated (e.g., Shlosman et al. 1989; Friedli \& Martinet 1993; Combes
1994; Maciejewski \& Sparke 2000; Heller et al. 2001; Shlosman \& Heller 2002; El-Zant \& 
Shlosman 2003), but modeling of nested bars in a cosmological scenario even for isolated
DM halos was never attempted or validated. 

This Letter demonstrates how a nested bar system forms in a disk growing 
in an assembling triaxial DM halo. We use cosmological initial conditions and 
follow the Hubble expansion and the subsequent collapse of an isolated perturbation in the 
gas and the DM. Our star formation (SF) algorithm is physically
motivated and we include feedback from stellar evolution. A large 
number of models have been run and show the formation of galactic disks whose 
structural parameters fit within the observed range. Here we 
describe a representative model only and focus on the formation of
nested bars. Additional aspects will be addressed elsewhere.

%%%%%%%%%%%%%%%%%%%%%%%%%%%%%%%%%%%%%%%%%%%%%%%%%%%%%%%%%%%%%%%%%%
\section{Numerics, Star Formation, Feedback and Initial Conditions}
%%%%%%%%%%%%%%%%%%%%%%%%%%%%%%%%%%%%%%%%%%%%%%%%%%%%%%%%%%%%%%%%%%

The simulations were performed with an updated version of the
FTM-4.4 hybrid $N$-body/SPH code (Heller \& Shlosman 1994),
using the routine {\tt falcON} (Dehnen 2002) to compute
the forces. The gas temperature was obtained from the energy equation.
We used $N_{\rm DM}=5\times 10^5$ and $N_{\rm SPH}=5\times 10^4$, 
gravitational softening of 150~pc for DM and stars, and dynamic 
softening with a minimum of 250~pc for the gas.
Tests have been performed to check the sensitivity of the
results to the algorithm and its parameters. A pure DM run 
conserved the energy to within 1\% and $J$ to within 0.1\%.

The modified prescription for SF and the feedback from stellar evolution 
take place in the Jeans unstable, contracting
region. We fix the gas-to-background density at which gas is converted
to a star, the local collapse-to-free fall time, and introduce the
probability that the gas particle produces a star during a given timestep.
Four generations of stars form per gas particle with an instantaneous    
recycling along with an increment in metallicity. 
The balance of the specific internal energy along with the gas 
ionization fractions of H and He and the mean molecular weight are 
computed as a function of $\rho$ and $T$ for an optically thin 
primordial composition gas.
Feedback from stellar evolution includes the supernovae and 
OB stellar winds, and uses the thermalization parameter
--- a fraction of the feedback deposited as a thermal energy, and converted
into kinetic energy via equations of motion (Heller et al., in prep.).  

The initial conditions are those of a spherically-symmetric density enhancement 
at $z=36$ evolved using an open CDM (OCDM) model with $\Omega_0=0.3$, $h=0.7$. The 
difference between the OCDM and $\Lambda$CDM is minimal on sub-galactic scales.
The initial density profile corresponds to the average density 
around a $2\sigma$ peak in a Gaussian random density field. The spin parameter 
$\lambda = 0.05$ is set by the angular 
velocity $\omega \propto r^{-1}$, where $r$ is the cylindrical radius and
the central kpc has been softened.
We impose the collapse redshift of $z=2$. The mass 
of collisionless DM particles is $7\times 10^{11}~{\rm M_\odot}$ and the gas
comprises 10\% of the total mass initially.

The main simplification in our models is that
we neglect interactions and minor or major mergers. Yet, because
the evolution is determined to a large extent by the nonlinear
interaction between three non-axisymmetric components, the results presented 
here can be considered representative.

%%%%%%%%%%%%%%%%%%%%%%%%%%%%%%%%%%%%%%%%%%%%%%%%%%%%%%%%%%%%%%%%%%%%%%%%%%
\section{Results}
%%%%%%%%%%%%%%%%%%%%%%%%%%%%%%%%%%%%%%%%%%%%%%%%%%%%%%%%%%%%%%%%%%%%%%%%%%

The initial perturbation expands with the Hubble flow and collapses, 
forming a DM halo --- a nearly {\it nonrotating} triaxial 
figure. In particular, the innermost 10~kpc experience a very abrupt
and short-lived increase of their ellipticity to
$\epsilon_{\rho}$ ~ 0.2, presumably due to radial orbit instability. The disk 
grows within the halo equatorial plane, visible already
at $\tau\sim 0.5$~Gyr. Its growth nearly washes out 
the inner, $<10$~kpc, halo ellipticity, due to the increased central mass concentration
in the system and the out-of-phase response by the disk. The inner 5~kpc
(10~kpc) of the DM become nearly axisymmetric and only slightly
flattened after $\sim 2$~Gyr ($\sim 7$~Gyr), while at larger radii $\epsilon_{\rho}\rightarrow 
0.15-0.25$ and $f_{\rho}\rightarrow 0.3-0.4$. 

\begin{figure}
\begin{center}
\psfig{figure=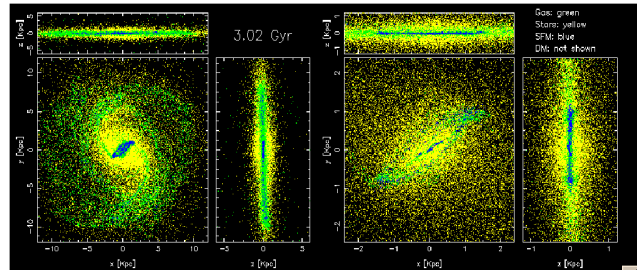,width=8.33cm,angle=0}
\end{center}
\caption{Dynamical evolution of nested bars --- a snapshot at $\tau\sim 3$~Gyr 
(see also the Animation Sequence~1). A large-scale bar (left frame) is shown in gas 
(green), stars (yellow) and SF (blue), and the nuclear bar in the same colors (right 
frame). The left frame is 24~kpc and the right frame is 4.8~kpc on the side. The 
Animation shows the evolution of the disk and the large-scale primary bar in the 
above colors (left frame) and of the inner kpc and the associated secondary nuclear 
bar (right frame) over 4~Gyr. 
}
\end{figure}

\begin{figure*}
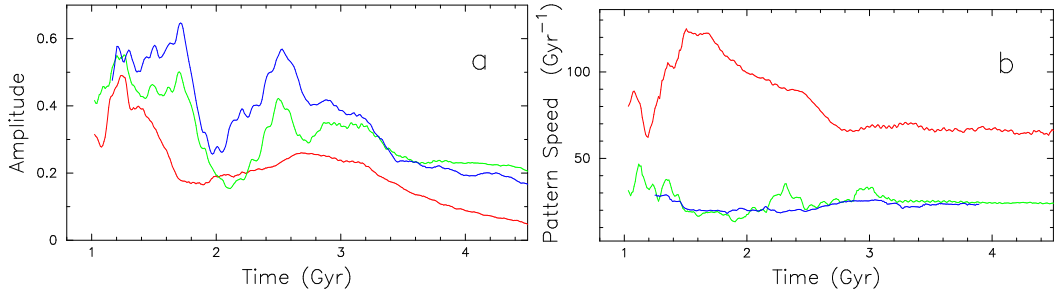

\begin{center}
\psfig{figure=fig2a.ps,width=3.8cm,angle=-90}\hspace{0.001cm}
\psfig{figure=fig2b.ps,width=3.8cm,angle=-90}\\
\end{center}
\caption{Nested bars evolution: $(a)$ $m=2$ amplitudes $A_2$, and $(b)$ associated 
pattern speeds. Shown are the stellar $A_2$ and pattern speeds of the secondary 
(red line) and primary (green line) bars and of the oval disk, between $5-10$~kpc (blue line). 
}
\end{figure*}

The disk evolution consists of gas-dominated (first $1.5-2$~Gyr) and star-dominated 
phases (Fig.~1 and Animation Seq.~1). The ratio of baryonic-to-DM matter within the 
central 10~kpc increases from $\sim 0.4$ to $\sim 0.54$ over the
Hubble time due to an adiabatic contraction.  
The corresponding pure DM model results in the NFW (Navarro et 
al. 1996) density profile and the characteristic radius $R_{\rm nfw}\sim 9$~kpc and is 
even more triaxial. For the model with baryons --- $R_{\rm nfw}\sim 4.5$~kpc, and the fit 
quality has worsened. The cusp is baryon-dominated. 

Initially, the disk is roughly oval and dominated in the outer parts by $m=2$ and 3  
grand-design spirals. Gradually it becomes more axisymmetric. The 
characteristic time for the disk (and halo) buildup, 
i.e., reaching 50\% of its mass at $z=0$, is $\sim 1.5$~Gyr. The rate of a baryonic 
inflow into the disk region crests 
within 1~Gyr. The SF rate shows a similar behavior. 

\begin{figure}
\begin{center}
\psfig{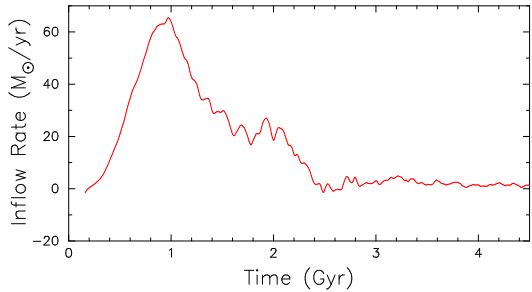}
\end{center}
\caption{Gas inflow rate across the central kpc. 
}
\end{figure}

The disk grows from inside out. A primary $\sim 2$~kpc-radius bar appears in the 
stellar component around 0.6~Gyr {\it normal} to the halo major axis, increasing to 3~kpc 
by 0.7~Gyr. By 0.8~Gyr, the inner part
of the 4~kpc bar, which is gas-dominated, decouples  
in the prograde direction and forms a secondary bar. By 1.5~Gyr, most of the gas in the secondary 
has fallen to the center --- further collapse is inhibited by the numerical resolution.

Fig.~2a follows the evolution after 1~Gyr. Both bars are strong at 
this time, as given  by $A_2\sim 0.6$ and 0.45 --- the amplitudes of corresponding 
$m=2$ modes. They are SF-dominated and involve a large fraction of the stellar and gas 
disk mass. The shape of the primary depends strongly on its orientation with respect 
to the halo {\it and} disk major axes --- main sources of tidal
torques --- and on
its internal (self-gravitational) response, e.g., between 1 to
2~Gyr (Anim. Seq.). The disk response to the halo torques drives a pair of strong 
grand-design arms whose
pitch angle gradually decreases forming a pseudo-ring outside the bar,
after which the open arms are regenerated (Anim. Seq.). The bars appear to
decay after $\sim 1.7$~Gyr in tandem with the arms, but strengthen abruptly after 
2~Gyr. The primary weakens again after 2.5~Gyr, while the secondary bar 
dissolves by $\sim 5$~Gyr. The 
bar sizes vary between $r_{\rm s}\sim 0.5-1$~kpc for the secondary and 
$r_{\rm p}\sim 5-7$~kpc for primary bars.  

The central issue of bars-in-bars systems is the {\it dynamical coupling} of bars as 
measured by the ratio of pattern speeds, $\Omega_{\rm s}/\Omega_{\rm p}$ (Fig.~2b). 
The primary bar slows down
prior to 1.5~Gyr, tumbles with a constant $\Omega_{\rm p}$ over the next
1~Gyr, speeds up and stabilizes around 2.8~Gyr for the rest of the simulation time.
The secondary bar speeds up dramatically before 1.5~Gyr, stabilizes for
$\sim 0.2$~Gyr, then slows down and stabilizes again after 2.8~Gyr. Its $\Omega_{\rm s}$
appears to have a plateau also around 2.4~Gyr. Hence the bars go over a number of
stages with various $\Omega_{\rm s}/\Omega_{\rm p}\approx const.$ 

Additional evidence for coupled evolution comes from varying corotation (CR) radii, 
$r_{\rm cr,s}$ and $r_{\rm cr,p}$, of bars and from the position of the inner Lindblad 
resonance (ILR) of the primary, $r_{\rm ILR,p}$. During the first 1.5~Gyr, $r_{\rm cr,s}$ 
and $r_{\rm cr,p}$ increase sharply from $\sim 1$~kpc and 4.2~kpc to 2.3~kpc and 8.2~kpc, 
respectively. This increase is caused by a substantial inflow that leads to a surge of 
the rotation curve. This is followed, between $1.5-2.8$~Gyr, 
by a plateau in $r_{\rm cr,s}(\tau)$ and $r_{\rm cr,p}(\tau)$, and a further increase 
of $r_{\rm cr,s}$ to 3.5~kpc. Both remain 
stable thereafter. We estimate $r_{\rm cr,p}/r_{\rm p}\sim 1.1-1.6$
and $r_{\rm cr,s}/r_{\rm s}\sim 1.3-5$. The value of 5 is achieved at the end of the 
first 1.5~Gyr (it results from the runaway action of the gas-dominated secondary, 
see \S 4), and after 2.8~Gyr --- making the secondary much shorter than its CR. 
The ILR of the primary can be inferred to lie outside the nuclear ring (see below) during 
this stage, $r_{\rm ILR,p}\sim 3.5$~kpc.

Nuclear rings form at the interface between the bars, and a pair of grand-design arms 
extend along the primary --- the ring and the arms 
are delineated by the SF intermittently. The ring evolution is 
more dramatic when viewed in the SF colors and it fades away with the
weakening of the primary.  
%The stellar bulge stands out after 2.5--3~Gyr. Its S\'ersic shape index
%tends to $\sim 1.3$.  

%%%%%%%%%%%%%%%%%%%%%%%%%%%%%%%%%%%%%%%%%%%%%%%%%%%%%%%%%%%%%%%%%%
\section{Discussion}
%%%%%%%%%%%%%%%%%%%%%%%%%%%%%%%%%%%%%%%%%%%%%%%%%%%%%%%%%%%%%%%%%%

We have simulated, in a large number of models, the formation of galactic disks 
in assembling triaxial DM halos and analyzed a representative 
model. The accompanying SF and the stellar feedback 
have been included. We followed the 
collapse of an isolated cosmological density perturbation with 
$\lambda=0.05$ from its linear regime, in the OCDM
universe. The disk and halo form, i.e., reach 50\% of their final mass, over a 
period of $\sim 1.5$~Gyr. The DM halo develops a clear
triaxial shape, elliptical in the plane perpendicular to the original
angular momentum axis and flattened along this axis. The resulting halo figure
tumbles very slowly, $\sim\pi$ over the Hubble time.

The halo triaxiality decreases during the disk growth because of the 
increasing central mass concentration and the disk response to the halo potential. 
The first effect reduces somewhat $f_{\rho}$ and $\epsilon_{\rho}$ of the halo,
even in the absence of a baryonic component. The second effect results
from the
negligible tumbling of the halo figure --- the inner ILR  
and the outer ILR are pushed to the center and to large radii, respectively. 
The disk responds out-of-phase with the halo potential, thus diluting the halo
ellipticity in its equatorial plane. This effect is not related to dissipation.

Most importantly, the halo torques lead to a strongly oval-shaped, growing disk 
and trigger the bar formation --- reminiscent 
of torques exerted on the disk during galaxy interactions. 
{\it These initial bars form as a result of a finite amplitude perturbation and not as a 
consequence of an exponential growth from an infinitesimal perturbation.} Their 
formation timescale is much shorter than for spontaneous bars and they appear
first normal to the halo major axis. In our simulation the initial
triaxiality of the inner halo is due to a short-lived episode of the
radial orbit instability erasing the initial conditions (e.g., MacMillan et al.
2006). Other causes of triaxiality are possible ---
we wish to stress here its effect on bar formation and evolution.

While nested bars have been simulated before, with various degrees of self-consistency,
this simulation differs in at least four major aspects. First, this 
is the most self-consistent model of nested bars in the literature --- no
{\it a~priori} assumptions have been made. Second, this is the first model where
nested bars form from cosmological initial conditions. Third, both bars are very 
gas rich, while previous
models of nested bars focused on purely stellar or gaseous bars (Shlosman 2005, for 
review). Fourth, we have incorporated SF and feedback  
--- the stellar and gaseous masses in the simulations are not
individually conserved. 

It is helpful to divide the nested bar evolution into three distinct phases, 
namely formation, dynamical adjustment and dynamically quiescent
phases, as  
exhibited in Fig.~2b. The {\it initial phase}, which is preceded by
an early disk, ends at $\sim 1.5$~Gyr and 
extends over $\sim 0.7$~Gyr from the time the secondary 
can be identified. The primary and especially secondary bars form gas rich. During 
this time, $\Omega_{\rm p}$ decays steadily  while $\Omega_{\rm s}$ increases 
monotonically by a factor of $\sim 2$. The latter increase is due to the gas collapse 
by a factor of 3--4 within the secondary, down to numerical resolution scales. 
Simulations of a dynamical runaway of a pure gaseous secondary have shown 
$\Omega_{\rm s}\sim r_{\rm s}^{-1}$ (Englmaier \& 
Shlosman 2004). A smaller growth in $\Omega_{\rm s}$ in our model can be explained
if a correction is made for the non-dissipative stellar component in the bar.
Note that $\Omega_{\rm p}$ follows closely the pattern speed of the oval disk, which 
forms in response to the halo ellipticity (Fig.~2b).
 
The {\it second phase} of evolution, between 1.5 -- 2.8~Gyr, is 
characterized by a dynamical adjustment of primary and secondary bars via their 
pattern speeds. The ratio $\Omega_{\rm s}/\Omega_{\rm p}\sim 6.1$ 
is steady for about 0.2~Gyr, then decays steadily because of a monotonic decrease in 
$\Omega_{\rm s}$, while $\Omega_{\rm p}$ stays constant initially, then increases 
slowly. The bars pass through an intermediate period of $\sim 0.2$~Gyr when their 
pattern speeds are possibly locked, $\Omega_{\rm s}/\Omega_{\rm p}\sim 4.4$, at about 
2.4~Gyr.
A dramatic inflow
leads to a buildup of a massive gas disk that generates grand-design shocks in 
response to the (elliptical) halo driving when the disk ellipticity surpasses a critical 
value (B.~Pichardo \& I.~Shlosman, 2005, unpublished). This process  largely determines 
$\Omega_{\rm p}$ which is also the pattern speed of the spiral arms. 
Hence, contrary to other models, the overall evolution in the system is driven by the 
halo triaxiality. 

While the disk shape depends strongly on 
the angle with the halo major axis, the primary bar is shaped by the
angle with the
halo {\it and} with the oval disk --- an efficient way to amplify chaos within the 
bar, weakening it on a short timescale (El-Zant \& Shlosman 2002; Berentzen et 
al. 2006; Berentzen \& Shlosman 2006). This can explain the sharp decay of primary 
amplitude at 1.7~Gyr and 2.5~Gyr, both accompanied by massive spiral arms
in the disk. Secondary bars have been found to adjust their properties, such as axial 
ratios and radial extent, depending on the mutual bar orientation (Heller et al.
2001; Shlosman \& Heller 2002). The situation is much more complex in the present run, 
because a number of {\it a}symmetric components coexist. The individual interactions, which 
are not always possible to disentangle, will increase the fraction of chaotic orbits
in the bars and generate local (non-grand-design) shocks in the gas. These 
interdependencies can have observational 
corollaries at redshifts corresponding to disk growth and will be addressed elsewhere. 
  
The {\it third phase,} after $\sim 2.8$~Gyr when the gas fraction is low, is 
that of a stable coupling between $\Omega_{\rm p}$ and $\Omega_{\rm s}$, i.e., 
$\Omega_{\rm s}/\Omega_{\rm p}\sim 2.7$, confirming that {\it stellar} nested bars can be 
long-lived systems (El-Zant \& Shlosman 2003). Because $r_{\rm ILR,p}\sim r_{\rm CR,s}$
at this time (see \S 3), it is possible that we observe a resonant coupling 
between primary and secondary 
as proposed by Tagger et al. (1987) for any two modes.
 
During the first stage, the gas inflow across the central kpc, $\dot M_1$, 
corresponds to a quasar-type activity of $60~{\rm M_\odot~yr^{-1}}$ (Fig.~3), 
while the SF peaks at $25~{\rm M_\odot~yr^{-1}}$. 
In the second stage, $\dot M_1$ crests at $\sim 25~{\rm M_\odot~yr^{-1}}$ 
around 2~Gyr, dropping to its lowest value close to zero thereafter. 
Nuclear rings and the associated
SF are intimately related to this inflow (Athanassoula 1992; Heller \& Shlosman 
1996; Knapen 2005), but in nested bars they vary in shape and have a more limited life 
span in response to the time-dependent potential (Shlosman \& Heller 2002).

Hence, we have demonstrated that nested bars in isolated halos form from 
cosmological initial
conditions and go through a series of dynamical couplings and decouplings, while
channeling the gas inwards to the smallest scales resolved numerically. This inflow can 
support the early quasar-type and Seyfert-type activity thereafter. The
exact conditions leading to an SBH formation (Begelman et al. 2006) are beyond the scope
of this work, but the remote `boundary' conditions are in agreement.

\acknowledgements
Support by NASA and NSF is gratefully acknowledged.


\begin{thebibliography}{}

\bibitem[]{420}Allgood, B., Flores, R.A., Primack, J.R., Kravtsov, A.V., Wechsler, R.H.,
       Faltenbacher, A., Bullock, J.S., 2006, MNRAS, 367, 1781

\bibitem[]{423}Athanassoula, E. 1984, Phys. Reports, 114, 320   
       
\bibitem[]{425}Athanassoula, E. 1992, MNRAS, 259, 345

\bibitem[]{427}Athanassoula, E. 2002, \apj, 569, L83  

\bibitem[]{429}Athanassoula, E. 2003, MNRAS, 341, 1179
   
\bibitem[]{431}Begelman, M.C., Volonteri, M., Rees, M.J. 2006, MNRAS, 370, 289 

\bibitem[]{433}Berentzen, I. Shlosman, I., Jogee, S. 2006, \apj, 637, 582 

\bibitem[]{435}Berentzen, I. Shlosman, I. 2006, \apj, 648, 807
   
\bibitem[]{437}Combes, F. 1994, in Mass-Transfer Induced Activity in Galaxies, (ed.) 
     I.Shlosman, CUP, 170

\bibitem[Dehnen (2002)]{Dehnen_02} Dehnen, W. 2002, J. Comp. Phys., 179, 27

%\bibitem[]{442}Elmegreen, B.G., Elmegreen, D.M., Hirst, A.C. 2004, \apj, 612, 191  
\bibitem[]{442}Elmegreen, B.G., et al. 2004, \apj, 612, 191 

\bibitem[]{444}El-Zant, A., Shlosman, I. 2002, \apj, 577, 626

\bibitem[]{446}El-Zant, A., Shlosman, I. 2003, \apj, 595, L41

\bibitem[]{448}Englmaier, P., Shlosman, I. 2004, \apj,  617, L115  
   
\bibitem[]{450}Erwin, P., Sparke, L.S. 2002, \aj, 124, 65

\bibitem[]{452}Ferrarese, L., Ford, H. 2005, Space Sci. Rev., 116, 523
   
\bibitem[]{454}Friedli, D., Martinet, L. 1993, A\&A, 277, 27

\bibitem[]{456}Grosbol, P., Patsis, P.A., Pompei, E. 2004, A\&A, 423, 849

\bibitem[Heller \& Shlosman (1994)]{Heller_Shlosman_94}
   Heller, C.H., Shlosman, I. 1994, \apj, 424, 84

\bibitem[]{461}Heller, C.H., Shlosman, I. 1996, \apj, 471, 143 

\bibitem[]{463}Heller, C.H., Shlosman, I., Englmaier, P. 2001, \apj, 553, 661

\bibitem[]{465}Jogee, S., Barazza, F.D., Rix, H.-W., Shlosman, I., et al. 2004,
    ApJ, 615, L105  
   
\bibitem[]{468}Knapen, J.H., Shlosman, I., Peletier, R.F. 2000, \apj, 529, 93

\bibitem[]{470}Knapen, J.H. 2005, A\&A, 429, 141

\bibitem[]{472}Laine, S., Shlosman, I., Knapen, J.H., Peletier, R.F. 2002, \apj, 567, 97       

\bibitem[]{474}Maciejewski, W., Sparke, L.S. 2000, MNRAS, 313, 745

\bibitem[]{}MacMillan, J.D., Widrow, L.M., Henriksen, R.N. 2006, astro-ph/0604418

\bibitem[]{476}Marinova, I., Jogee, S. 2006, astro-ph/0608039

\bibitem[]{478}Navarro, J.F., Frenk, C.S., White, S.D.M. 1996, \apj, 462, 563 (NFW)   

\bibitem[]{480}Sellwood, J.A. 2006, \apj, 637, 567

\bibitem[]{482}Sheth, K., Regan, M.W., Scoville, N.Z., Strubbe, L.E. 2003,
    \apj, 592, L13    

\bibitem[]{485}Shlosman, I., Frank, J., Begelman, M.C., 1989, Nature, 338, 45

\bibitem[]{487}Shlosman, I., Begelman, M.C., Frank, J. 1990, Nature, 345, 679
    
\bibitem[]{489}Shlosman, I., Heller, C.H. 2002, \apj, 565, 921

\bibitem[]{491}Shlosman, I. 2005, in The Evolution of Starbursts, S. Huettemeister \& 
     E. Manthey (eds.), Melville: AIP, p. 223

\bibitem[]{494}Tagger, M., Sygnet, J.F., Athanassoula, E., Pellat, R. 1987, ApJ, 318, L43

\bibitem[]{496}Weinberg, M. D. 1985, MNRAS, 213, 451

\end{thebibliography}
\end{document}